# AN ETHICAL STUDY OF GENERATIVE AI FROM THE ACTOR-NETWORK THEORY PERSPECTIVE

Yuying li And Jinchi zhu

Department of Journalism and Communication, Communication University of Zhejiang, Hangzhou, China

## ABSTRACT

*The widespread use of Generative Artificial Intelligence in the innovation and generation of communication content is mainly due to its exceptional creative ability, operational efficiency, and compatibility with diverse industries. Nevertheless, this has also sparked ethical problems, such as unauthorized access to data, biased decision-making by algorithms, and criminal use of generated content. In order to tackle the security vulnerabilities linked to Generative Artificial Intelligence, we analyze ChatGPT as a case study within the framework of Actor-Network Theory. We have discovered a total of nine actors, including both human and non-human creatures. We examine the actors and processes of translation involved in the ethical issues related to ChatGPT and analyze the key players involved in the emergence of moral issues. The objective is to explore the origins of the ethical issues that arise with Generative Artificial Intelligence and provide a particular perspective on the governance of Generative Artificial Intelligence.*

## KEYWORDS

*Generative Artificial Intelligence , Actor-Network Theory , ChatGPT , Artificial Intelligence Ethics*

## 1. INTRODUCTION

Generative artificial intelligence (GenAI) encompasses a diverse array of tools, and its development signifies a paradigm shift in the application of machine learning technology. GenAI empowers machines to transition from developing tools for identifying hidden patterns in text, photos, or other disorganized datasets to producing unrestricted text, images, and video using algorithms that have been trained on extensive datasets. [1] Currently, GenAI exhibits promising potential in various domains, including medicine, the arts, education, and finance. Chui et al. [2] conducted a recent global survey to assess the adoption of AI among organizations. The survey included 1,684 respondents. The findings revealed that approximately one-third of the respondents reported regular usage of GenAI for at least one function within their organizations. Additionally, 40 percent of the respondents expressed their companies' interest in increasing investments in AI, specifically GenAI. Furthermore, 28 percent of the respondents mentioned that the use of GenAI was already being considered by their organizations' boards. It is evident that GenAI is significantly impacting different businesses, particularly the content production industry. Simultaneously, the study report revealed that only a small number of organizations were sufficiently equipped for the extensive use of AI or the potential commercial hazards associated with these technologies. Merely 21 percent of participants stated that their firm had implemented a policy. When directly questioned about the potential hazards associated with utilizing AI, only a small number of respondents acknowledged the company's preparedness to address the uncertainties of GenAI. Potential risks that GenAI may have are also being disclosed, with privacy violations, data breaches, algorithmic discrimination, copyright attribution, and other issues popping up all over the place.

The ethical concerns associated with GenAI have expanded due to its broad implementation and technological advancements. In the social field, GenAI's ethical issues extend from privacy risks to escalating social inequality [3]. In the realm of education, it raises questions about academic quality [4], fairness[5], the impact of bug fixes on efficiency, and the replicability of AI research [1]. In the field of information technology, GenAI's technological vulnerability presents opportunities for cybercrime [6]. Meanwhile, Illia et al. (7) introduced three new ethical challenges that could hinder the success of GenAI texts by combining agenda-setting theory and stakeholder theory. These challenges include: (1) the potential for mass manipulation and dissemination of false information; (2) the possibility of mass producing low-quality yet believable content; and (3) a notable decrease in direct communication among stakeholders. To address these issues, scholars endeavor to discover novel answers and breakthroughs by using technology itself, employing other technological advancements, and employing creative research methodologies. Gupta et al. [6] employed GenAI techniques to enhance security mechanisms with the aim of mitigating cybercrime resulting from abuse. Chimbga [8] examines the ethical and logical incorporation of stakeholders in the IoT ecosystem; Michel-Villarreal et al. [3] This study creatively investigates the possible benefits and constraints of ChatGPT by conducting in-depth interviews with ChatGPT in a semi-structured manner. By engaging in a thorough examination of these matters, scholars aim to gain a comprehensive understanding of the incorporation of technology from a fair and reflective standpoint [5]. They also seek to effectively utilize GenAI by tackling these challenges and harnessing the capabilities of technology in order to bridge the divide between the intricate intricacies of technology and human ingenuity [9].

Based on current research, the majority of experts have extensively examined the issues surrounding GenAI across several disciplines. Therefore there is a need for a flat and systematic study of the reasons behind it. Numerous factors influence the ethical conundrums surrounding GenAI, necessitating a thorough and objective analysis of their crucial factors. This research uses actor network theory to examine and evaluate ChatGPT, a prominent GenAI, as a specific instance. This research utilizes Actor-Network Theory, which examines the action network of ChatGPT as a flat society where actors are assumed to have equal status, enhancing the understanding of translation between the actors. This research aims to analyze key actors that impact the rise of ethical dilemmas to delve further into their origins. This study suggests the following research questions to tackle this issue:

- What is the structure of ChatGPT's network of actors? What is its composition?
- Which actors and translation process in the network are responsible for the emergence of ChatGPT ethical issues?
- Which key actors impact the emergence of ethical issues?

In the second portion, we will construct the actor network of ChatGPT and justify the use of ChatGPT as a case study for generative AI in the initial phase of this research. In this portion, we will discuss the rationale behind selecting and defining the actors, as well as the scientific process of constructing the network. The network framework will be provided for the case study in the subsequent section. The third part will explain how actors in the ChatGPT network result in moral issues through translation. Four ethical issues in ChatGPT include algorithmic bias, illegal application, trust collapse, and privacy violations. We consider these four difficulties as dimensions and utilize the ChatGPT actor network to study the actors participating in translation in each dimension and identify the key actors. This paper integrates the technical aspects of ChatGPT with an examination of the transition from the laboratory to the market and ultimately to the platform. It also scrutinizes the key actors involved in ethical concerns to elucidate the underlying causes of ethical issues arising in the utilization of ChatGPT.

## 2. CONSTRUCTING A NETWORK OF CHATGPT ACTORS

French sociologist Bruno Latour developed Actor-Network Theory in the 1980s as a paradigm for sociological and technological study. The fundamental concept of Actor-Network Theory is to regard both human and non-human actors as equal participants within a network and to eliminate the hierarchical structure of society by means of network analysis [10]. According to Actor-Network Theory, the process of connectivity and collaboration between actors relies on the establishment of an "obligatory passage point" (OPP) by influential individuals in the network. Consensus among actors is necessary before any action can be successfully carried out. This action is translation. Translation facilitates the connection and collaboration of individuals within a particular setting, hence establishing a social network.

### 2.1. Establishment Of The ChatGPT Network Of Participants

The ethical issues surrounding GenAI are extensive, and a specific case study is necessary to thoroughly evaluate the underlying reasons. Given that GenAI is an emerging technology for generating content based on models, it is worthwhile to examine the widely used ChatGPT as a case study. This analysis allows us to gain insights into the underlying principles and production process, enabling us to understand the finer details while also grasping the broader context. Ultimately, this examination will provide valuable reference experiences for exploring the ethical considerations surrounding GenAI. OpenAI developed ChatGPT, an artificial language model. It works as a chatbot and incorporates natural language processing capabilities. The ChatGPT industry chain encompasses several aspects such as technology development, marketing, consumer adoption, market penetration, and technological review and iteration. These components are interconnected and rely on three primary pillars: technology, human resources, and capital. Actor-Network Theory requires the careful examination and analysis of actors, their translations, and the intricate relationships within the ChatGPT actor network. This is essential for gaining a comprehensive understanding of the interplay of interests resulting from translation, the subsequent impact on actors, and the emergence of ethical dilemmas.

Technological advancements frequently arise in response to human demands, serving as the fundamental basis and inherent component of social behaviors. [12]. When the necessity is consistently reinforced, it may transform into a demand. Demand is an elusive and latent awareness. Freud compared consciousness to "the visible part of an iceberg floating on the water's surface". The model developers have an underlying desire for this hidden demand to manifest itself in technology in order to fulfill the user's requirements and entice them to purchase and utilize it. User feedback and emerging demands drive the iterative development of technology, with the user's part being integral to all elements of its emergence. Thus, inside the ChatGPT network, users play a central role, and the objectives they establish to meet their needs are interconnected with the relevant participants in the network. The advancement of ChatGPT technology is driven by consumers' demand for enhanced capabilities. ChatGPT is a model that requires dataset training to enhance its accuracy and develop a fundamental comprehension of the data. The training process of the dataset involves two distinct processes, with the initial step being the pre-training of the dataset. The pre-training dataset is obtained from a data collection provider that subjects the model to initial training using a vast dataset. The second dataset is the fine-tuning dataset, which involves using meticulously labeled data of superior quality to train the model for a particular task. The integration of these three components brings about the implementation of ChatGPT technology. Following the availability of the technology, producers of AI tools modified and bundled the model into various formats, such as web pages, applications, and program software. This was done to cater to the specific requirements of professionals in different industries. Following the release of OpenAI Inc.'s API for the ChatGPT 3.5 model, a plethora of ancillary apps surfaced, including ChatPDF, ChatExcel, ChatBCG, and others. Users of AI employing ChatGPT models and associated tools will inadvertently contribute to the market expansion of ChatGPT. This will occur through both robust and feeble connections between media platforms and offline channels, potentially influencing users to take notice of and embrace ChatGPT. Consequently, users of AI will become a significant player within the network of ChatGPT actors.

Technology plays a significant role in the ChatGPT actor network; therefore, a detailed analysis will be conducted on the non-human players inside the network. This analysis will involve deconstructing the process of model generation and examining the essential actors involved. Based on the Principle of General Symmetry and Inscription of actor network theory, the algorithms and personnel responsible for media platforms and data gathering firms will not undergo refinement, and both will be classified as non-human actors.AI users are examined as distinct entities within the user group by their more significant function in the network. Thus, we extracted the AI user from the user as an independent actor for the study. Human actors possess greater intricacy compared to non-human actors. For instance, AI users can further be categorized into professional and non-professional users. The question of whether they exhibit a willingness to share during the usage process is a subject of debate. To ensure proximity to the research topic and facilitate the construction of ChatGPT's network framework, we will streamline the study of human actors. Specifically, we will identify and analyze human actors who engage in multiple translation behaviors and possess substantial influence within the actor network. This paper contends that the primary participants in the ChatGPT actor network comprise pre-training datasets, fine-tuned training datasets, generative AI models, non-human actors affiliated with data collection companies and media platforms, and human actors encompassing generalized network users, AI users, model developers, and AI tool providers.

## 2.2. The ChatGPT Actor Theory Framework

The primary component of activity inside the ChatGPT actor network is distinguished by its diversity and heterogeneity. While actors demands are central to the action, many actors within the network have distinct objectives that directly influence their mobility. This results in them assuming diverse roles, and the interplay between these actors collectively shapes the ChatGPT action network. The ChatGPT actor network is centered around the interconnection of actors, facilitating the transfer and exchange of information across the nine actors, hence fostering the integration, advancement, and transformation of ChatGPT. Simultaneously, the malfunctioning of the translation process has resulted in numerous ethical dilemmas. Through the examination of the translation process employed by actors , one may effectively investigate and condense the ethical dilemmas that may emerge in many facets of ChatGPT. (Figure 1)

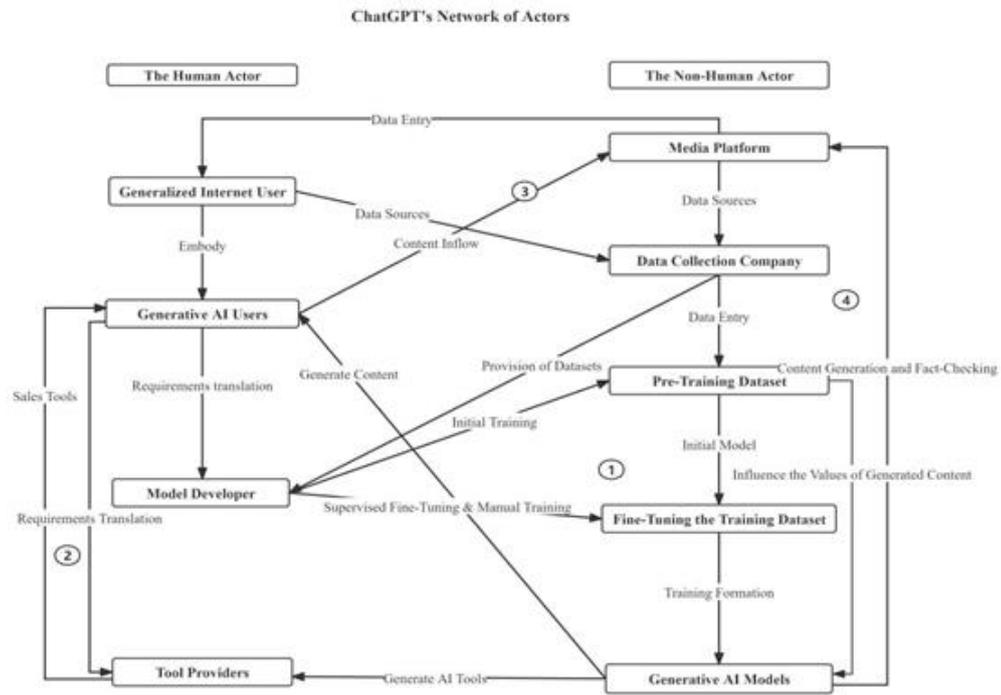

Figure 1. Framework for analyzing the network of generative AI actors

## 3. ETHICAL ISSUES ARISING IN CHATGPT AND THEIR REASONS

Translation facilitates the connection and collaboration among participants in a network. Without translation, the network would cease to exist. The actors inside the network carry out the process of translation, which leads to the modification of identity and attributes. This translation process enables them to exert control over other members and creates new power dynamics [14]. The presence of translation gives rise to a multitude of ethical concerns within the ChatGPT network. Hence, it is imperative to elucidate the process of translating and being translated among participants in the network in order to ascertain the cause of the emergence of ethical issues.

### 3.1. Quantity-Value Imbalance: Algorithmic Bias Derived from Data Sets

Two training processes—dataset completion and model translation—determine the final performance of ChatGPT. These processes involve the data collection company, the model developer, and multiple translation steps. It is important to note that algorithmic bias can be generated during these processes. The origin of this bias stems from the disparity in control over development cost and social value between the dataset-gathering firm and the model developer.

The original Transformer model [15] used in ChatGPT was not specifically designed to create an intelligent dialogue system. Its primary function is to predict missing words in a sentence. The developer does not have the ability to introduce algorithmic bias by manipulating certain parameters, which is challenging to achieve at the fundamental level of ChatGPT. When examining the model, we can liken it to a newborn baby. Similar to a baby, a newly generated model possesses the capacity to acquire knowledge but lacks any pre-existing knowledge. Its comprehension of the universe is derived from acquiring knowledge. In the pre-training phase, the ChatGPT model undergoes a data translation process. This process involves selecting a sentence from the dataset and randomly masking a word or subword within that sentence. The model then predicts the masked word by considering the context and identifying the most likely candidate word. The masked word in the original sentence serves as a supervised signal, allowing for a comparison between the model's prediction and the actual word. This comparison is used to calculate the loss, and the error signal is back-propagated to update the model parameters. This iterative process aims to enhance the model's performance in predicting words. The model developer accomplishes the process of translating data into model pre-training by utilizing the pre-training of the data provided by the data collection firm. This involves extracting the embedded features, patterns, and knowledge from the data and training them into the model. The data's value significantly influences the value of the model output's content. During the translation of data into the model, the data provided by the data collection firm becomes a crucial factor in the emergence of the algorithmic bias problem.

The demand for dataset development in the ChatGPT network originates from model development companies, such as OpenAI. The pre-training corpus for ChatGPT is sourced from various datasets, such as the CommonCrawl dataset, Reddit links, books and journals, and English Wikipedia [16]. These datasets have been collected and published by different organizations, including Zhuetal (2015), Shawn Presser, the Wikimedia Foundation, and the Common Crawl team [17]. Model creators require data solely to confirm the accuracy of the semantic level during training. Simultaneously, model developers want extensive data translation capabilities, typically necessitating thousands of gigabytes of data. This entails sourcing substantial amounts of data from multiple data-gathering businesses to meet the requirements. Hence, data collection companies frequently prioritize the quantity of data sets and their semantic accuracy, disregarding the inherent value and social consequences of the information. This approach can lead to the inclusion of "harmful" information, thereby introducing biased content into the model's learning process from the outset. While fine-tuning training developers have the opportunity to alter the likelihood of generating outcomes by manually manipulating data and implementing reward and punishment mechanisms, these actions have minimal effect on the pre-training dataset. Hence, algorithmic bias arises due to the influence of biased content in the dataset on the values of the model output during the translation process involving the data collection firm, the model developer, and the pre-training dataset.

The primary approaches employed in ChatGPT's fine-tuning training involve supervised automatic fine-tuning with human-authored data and manual reward and punishment training of the model utilizing reinforcement learning techniques based on human feedback. Both methods of fine-tuning training involve the notable influence of subjective human factors. The actors in this session, guided by professionalism and human preference, translate the model twice and remove any content from the initial model that does not align with human language conventions and ethical standards. OpenAI has a limited number of experimenters to fine-tune the model creation throughout the training phase of the fine-tuning process in order to manage the expense of development. The model developers enlisted the services of 40 annotators from two companies to manually generate a fine-tuned corpus for ChatGPT during the fine-tuning training process [18]. Despite their efforts, the 40 annotators were unable to entirely eliminate the biased data that the model had already acquired, in contrast to the vast number of pre-training datasets. Simultaneously, the fine-tuning training necessitated a subjective intervention to train a human preference reward model, which would penalize the generation of "harmful" speech by ChatGPT [19]. During this procedure, the experimenter has the potential to manipulate the model by introducing biased input that is influenced by their subjective impressions. In the case of ChatGPT, OpenAI, the model developer, conducted the fine-tuning training in the GenAI domain. However, there are other situations in which the technology provider creates the basic structure of the model while the technology buyers, such as AI tool providers and AI users, carry out the fine-tuning task in accordance with their particular requirements. It is quite probable that algorithmic bias will arise due to factors such as data gathering capability, technology, values, etc. This bias may also facilitate the development of generative AI applications in illicit or unethical domains.

### 3.2. The Collusion Between Tools and Content: The Unauthorized Utilization of Gray Market Demand.

While ChatGPT possesses impressive capabilities, it currently lacks the ability to autonomously generate content suitable for direct use. Hence, the dissemination of AI-generated material necessitates translation by both the AI tool supplier and the AI user prior to its market penetration..OpenAI offers two methods for utilizing ChatGPT on its official website: one involves engaging with it via the provided chat interface, while the other entails employing it through APIs. AI tool providers develop generative AI tools using API interfaces to facilitate the translation of ChatGPT from models to tools, thereby assisting AI users in overcoming technical obstacles. The translation of ChatGPT involves the transition of the model into a tool that aids AI users in overcoming complex obstacles. AI tool suppliers typically develop products that align with the requirements of consumers. Due to the nascent state of generative AI, the norms governing its use are not yet optimal, and the individuals involved are highly likely to influence the translation process with ethical dilemmas.

While the majority of AI users select AI tools based on genuine functional requirements, there is a risk that individuals involved in illicit activities in the gray and black industries may exploit the capabilities of ChatGPT. This is primarily due to the inconsistency among tool providers. Consequently, there is a high probability that this demand will be channeled towards generative AI tools, resulting in the misuse of ChatGPT for the creation of illegal content. Such applications are highly focused, leveraging both the robust content generation capabilities of GenAI and the accumulated expertise of content providers, and can significantly harm society. The modeling company aims to open up the API to enable different sectors to effectively utilize GenAI's features for industry development and generate income for the company. However, APIs have simultaneously generated technical vulnerabilities and a lack of regulation, which has allowed the gray sector to exploit advanced technology for profit. If you request illicit information from ChatGPT or ask it to fabricate information, it is highly likely that your request will be denied. Nevertheless, by leveraging the API interface to access ChatGPT, unethical technicians might circumvent regulations through several methods, hence allowing ChatGPT to generate unrestricted content.

To summarize, it is important to note that utilizing AI tools for content production in illicit industries involves more than just basic knowledge of generative AI technologies. It necessitates a certain level of technical expertise. The requirement for translation between AI users and tool providers is what surpasses the barrier between content and technology. During the translation process, there are distinct roles played by AI tool providers and AI users within the generative AI network. AI users express their demands and create a market for AI tools, while AI tool providers offer these tools to users in order to generate economic gains. Through mutual agreement, both parties achieve a shared interest, resulting in successful translation between them. Nevertheless, the intricate nature of the individuals involved in the translation process leads to the estrangement of the demand, model-to-tool, and tool-to-content translation processes.

### 3.3. The Disparity Between Content Production and Platform Regulation: AIGC Pandemic Leads to Confidence Collapse

The advent of Web 3.0 has occurred following multiple technological advancements. Currently, the advancement of generative artificial intelligence technology has eliminated the stereotyped and inflexible nature of AIGC content. Its enhanced output capacity and comprehension of abstract concepts can greatly assist AI users in creating content. The ChatGPT model assists AI users in efficiently resolving issues as it caters to their specific content requirements, hence significantly contributing to its user adoption and promotion.

The utilization of AIGC in media platforms is evident not only in the generation of content by AI users but also in the production of media material and verification of facts, hence facilitating the translation from model to content. During the process of translation, the inherent constraints of the individuals involved may give rise to the creation of inaccurate information. The concept has two notable limitations: the production of misinformation and the inability to dynamically update the knowledge base. The production of disinformation in ChatGPT is a result of the fundamental logic of its basic technology, neural networks. Neural networks primarily excel at predicting unknowns using available data. The ChatGPT model acquires the compositional characteristics of a text by converting it into mathematical vectors. Through extensive data training, the model becomes capable of probabilistically predicting its composition. By extensively analyzing data, the model is capable of generating content based on probabilities. However, the model lacks the ability to comprehend the meaning of the text, resulting in what is commonly referred to as "serious nonsense" responses. Overcoming this issue within a short timeframe is challenging, given the current state of technology. Regarding the second problem, the existing advanced AI technology often necessitates dual training. However, each training session demands a substantial allocation of computational resources and time, preventing the model from updating its knowledge base in real-time.The knowledge repository for GPT-4 is limited to information available until April 2023. This means that the pre-training dataset used for GPT-4 consists of information from 2023 and earlier. When a user requests information beyond that timeframe, GPT-4 can only provide answers and make guesses based on the information provided in the ongoing conversation. It cannot use this new information as a parameter to update its training data model.While GPT-4 possesses the capability to search for answers on the internet, it lacks the ability to translate new information from data to model and include it internally. Therefore, the network answers provided by GPT-4 are equivalent to those obtained from a browser search for an AI user.

The excessive reliance on AIGC by media platforms without adequate content regulation poses a significant risk to the authenticity of ChatGPT's content translation. This, in turn, can greatly undermine the social trust mechanism and give rise to issues concerning content authenticity. As AIGC technology advances, it is increasingly being integrated into media platforms for tasks such as content editing, organizing, and review, ultimately delivering the information to viewers. During the translation process, five actors play a role: models, AI users, AI tool providers, media platforms, and network users. It is possible for AI users to disseminate inaccurate information through the media while translating content. Furthermore, media platforms have witnessed the emergence of advanced forgery techniques that utilize generative AI technology. For instance, social media bots that appear on media platforms have the capability to disseminate content either through third-party platforms or through clients using automated scripts [20]. This is a formidable obstacle for censorship. Multiple studies have demonstrated that the identification methods for social bots exhibit both "false-positive" and "false-negative" inaccuracies. Additionally, there are instances where social bots are incorrectly classified [21]. Simultaneously, the media can employ the model to produce erroneous text and images, resulting in a substantial volume of AIGC content that is indistinguishable from reality appearing on media platforms. This poses a challenge for platforms to ensure the accuracy of the information they provide, potentially leading to the Tacitus effect and causing a significant crisis of trust in society.

### 3.4. Conflict Between User Rights and Data Usage: Privacy Violations Resulting from Information Collection

The utilization of Artificial Intelligence technology is contingent upon the support of data and model translation. Furthermore, a model that lacks its learning material will be incapable of making independent predictions. This holds particularly true for the ChatGPT model, which relies on an extensive pre-training dataset. However, in the current state of the field, one of the primary methods for a data collection business to acquire a substantial volume of data is by gathering user-generated content from the Internet. The user-generated material in the pre-training dataset of ChatGPT amounts to petabytes of data, surpassing the fraction of corpus content from books and journals. This discrepancy is easily comprehensible. The usefulness of ChatGPT is easily comprehensible. A large amount of daily communication data is needed to give ChatGPT features like dialog interaction and sentiment analysis that are similar to human communication. Social media platforms provide the most suitable source of data that fulfills these requirements. When turning web content into datasets, the issue of user data violations eventually emerges.

One type of content found on digital platforms is user-generated content. The users of ChatGPT are the primary participants in the entire web ecosystem. The primary rationale for its designation as the starting point is the substantial reliance on user-generated material from diverse websites, media outlets, and platforms for training AI models. The majority of the data sets in the pre-training dataset of ChatGPT, namely the Common Crawl dataset and the Reddit Content Aggregation Class dataset, are sourced from the Internet. The primary source of UGC content production consists of the platform users. The primary source of UGC (user-generated content) generation is the users of the site. Both datasets are sourced from the Internet and consist of user-generated web material. The primary source of UGC material production is the platform's users, and strictly speaking, only the users should possess the entitlement to utilize it. Nevertheless, currently, the identification of UGC content is indistinct. Data collection companies can establish direct agreements with social media platforms to gather user-generated content (UGC) data. These companies, along with media platforms, form a network of actors with shared interests. In situations where users are unaware, their information can be included in the data set without their knowledge. Additionally, network crawlers and other technical methods can be used to bypass regulations and directly extract content from websites. The authors of a newly published work on the latest multimodal model, Clip, state that their team utilized 400 million image-text pairs sourced from the Internet for training purposes [22]. This method of gathering data jeopardizes the rights of platform users and raises clear concerns regarding data ethics.

The second aspect pertains to the aggregation of users' confidential data. The dataset sources for generative AI heavily rely on user-produced information. When compared to user-generated content (UGC) data, user-private information serves as an even more valuable source of high-quality data. Instances of people engaging in conversations on instant messaging platforms like WeChat and WhatsApp, as well as private conversations on platforms like Twitter and Instagram, and even photographic photos stored on users' mobile devices, might serve as training data for models. While the training dataset of GPT3 that OpenAI has publicly released does not contain such information, there are still numerous experts who posit that this type of data has been utilized in training huge models. The Clarkson Law Firm, based in California, has alleged that OpenAI engaged in covert data collection by crawling around 300 billion words of online content, encompassing books, articles, websites, posts, and personal information, without obtaining consent. This alleged action has affected millions of users, according to the law firm. This data collection behavior extends to the users and has led to a significant infringement on their privacy.

We believe that the primary cause of algorithmic bias is the model developer's focus on controlling development costs. The ethical issue involves five actors: model developer, data collection company, pre-training dataset, fine-tuning training dataset, and model, as well as four translation processes: model developer and data collection company, model developer and pre-training dataset, model developer and fine-tuning training dataset, and fine-tuning training dataset and model. The primary reason for illegal use of GenAI is the exploitation of APIs by AI tool suppliers. The issue stems from the fact that tool providers interpret AI user requirements for specific gray and black industries, and models offer open APIs, providing opportunities for tool suppliers. The ethical issue of illegal use comprises three participants and two translation procedures. The primary reason for the decline in public trust in AIGC is the improper use of AI technologies by media platforms and the spread of inaccurate AI content by these platforms. There are five actors—models, AI users, network-wide users, and tool providers—engaged in four translational behaviors within media platforms. The main problem with privacy breaches is the reasonableness with which data collection companies gather datasets. Data collection firms obtain datasets by reaching a consensus with media platforms and using technology to access user-generated content and private information. Hence, the issue of privacy infringement includes three actors and two translation procedures.

## 4. CONCLUSION

According to the perspective of Actor-Network Theory, the ChatGPT network consists of both non-human actors and human actors, who are the nodes of the network. These nodes are interconnected to facilitate translation, creating a heterogeneous network coalition. The analysis of the ChatGPT actor network suggests that it is a closed-loop system with interconnected actors linked through direct or indirect relationships, maintaining a dynamic equilibrium. The performers' strong connections make them  highly susceptible to influencing the translation process. The heterogeneity of the actors involved and the relevance of their interests in ChatGPT result in the emergence of moral issue. The key actors affecting the emergence of ethical issues such as algorithmic bias, illicit use of GenAI, reduction in public trust, and privacy breaches are model developers, AI tool providers, media platforms, and data collection companies, respectively. While we discussed key actors that impact the development of ethical dilemmas, it is crucial to note that the ethical concerns surrounding ChatGPT do not stem from a singular actor. Our definition of a significant actor is one that serves as a node where multiple translation processes act, and ethical issues would not have surfaced without the participation and collaboration of other actors. Every ethical issue includes a minimum of three actors and arises from undergoing multiple translation.

The study's limitations are vital to acknowledge. The development company controls the internal information of ChatGPT's model and the training dataset, so we don't have access to them. This limits our ability to refine technical details or analyze the dataset further. In addition,  the actors in the ethical problem are mixed with individual factors that are difficult to define. For example, it is challenging for us to determine which AI users require illicit applications. This study encompasses the entire process from developing ChatGPT in the laboratory to bringing it to the market, so we selected the main actors in the network. Future research can focus on examining actors and their translations in relation to a specific procedure or a new ethical problem.

In conclusion, this article analyzes the translation between non-human actors (pre-training dataset, models, fine-tuning the training dataset, data collection company, media platform) and human actors (generalized Internet users, generative AI users, model developers, tool providers). Algorithmic bias, illicit use of ChatGPT, reduction in public trust, and privacy breaches are ethical issues. Model developers, tool providers, media platforms, and data collection companies are essential actors in this ethical dilemma. However, we discovered that ethical issues arise when key actors engage in mutual translation with other relevant actors. In the future, while regulating the ethical concerns related to GenAI, it is crucial to identify key influences and provide equal consideration to all parties involved. By strategically organizing and distributing power among different actors, we may foster the positive and structured advancement of generative AI.

## AUTHORS

YUYING LI  received the B.S degree from Shijiazhuang University in 2022, and now she is currently studying for a master's degree at the Communication University of Zhejiang .Her current research interests include journalism and communication.

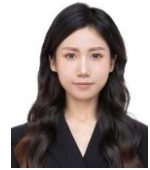

Jinchi Zhu received the B.S degree from Communication University of Zhejiang in 2022, and now he is currently studying for a master's degree at the same university. His current research interest is image quality evaluation in artificial intelligence

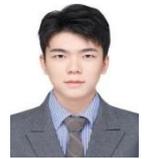